\documentstyle[preprint,aps]{revtex}
\tightenlines
\begin{document}
\draft
\title{Absence of a Bulk Meissner State in RuSr$_{2}$GdCu$_{2}$O$_{8}$}
\author{C. W. Chu$^{1,2}$, Y. Y. Xue$^{1}$, R. L. Meng$^{1}$, 
J. Cmaidalka$^{1}$, L. M. Dezaneti$^{1}$, Y. S. Wang$^{1}$, \\B. Lorenz$^{1}$ 
and A. K. Heilman$^{1}$}
\address{$^{1}$Department of Physics and the Texas Center for Superconductivity, 
University of Houston, Houston, TX 77204-5932\\
$^{2}$Lawrence Berkeley National Laboratory, 1 Cyclotron Road, Berkeley, 
CA 94720}
\date{\today}
\maketitle
\begin{abstract}
We have systematically investigated the magnetic, electrical, and structural 
properties of RuSr$_{2}$GdCu$_{2}$O$_{8}$, in which a long-range ferromagnetic 
order and superconductivity have been previously reported to coexist. 
Based on the reversible magnetization results, we conclude that the bulk 
Meissner state does not exist in 
this compound and that the condensation energy associated with 
superconductivity is 
negligible. The absence of a bulk Meissner state and the 
superconductivity detected are thus attributed to the possible appearance 
of a sponge-like 
crypto-superconducting fine structure in 
RuSr$_{2}$GdCu$_{2}$O$_{8}$ samples that are found to be chemically 
homogeneous to 1--2 $\mu$m and electrically uniform to $\sim$~10~$\mu$m across the 
sample.
\end{abstract}

The general antagonistic nature between superconductivity and magnetism 
has been long recognized. Over the last two decades, extensive studies 
\cite{maple82} have 
been carried out on a class of compounds known as ferromagnetic superconductors, 
in which the superconducting transition temperature ($T_{s}$) is higher than 
the magnetic transition temperature ($T_{m}$). While antiferromagnetism 
coexists with superconductivity, ferromagnetism does not, except in cases 
where the ferromagnetic order modifies its structure to a spiral or domain-like 
form to fit 
the superconducting state \cite{anderson59,ferrel79}. Until very 
recently, 
the coexistence of ferromagnetism and superconductivity has been observed to 
occur only below $T_{m}$ or only over a 
very narrow temperature region below $T_{m}$. However, a new class of compounds, 
called superconducting ferromagnets \cite{sonin98}, with a higher $T_{m}$ than 
$T_{s}$ was 
recently reported to show the coexistence of superconductivity and a uniform 
ferromagnetic order below $T_{s}$. They are layered ruthenate-cuprates 
RuSr$_{2}$GdCu$_{2}$O$_{8}$ (Ru-1212) \cite{bernhard99} and 
RuSr$_{2}$(RE$_{0.7}$Ce$_{0.3}$)$_{2}$Cu$_{2}$O$_{10}$ (Ru-1222) with RE = Gd 
or Eu \cite{felner97}. Their transition temperatures ($T_{m}$, $T_{s}$) are 
(133 K, 14--47 K) for Ru-1212, (180 K, 42 K) for Ru-1222 with RE = Gd, and 
(122 K, 32 K) for Ru-1222 with RE = Eu. Therefore, the observation is indeed very unusual and 
extremely interesting. It is particularly so in view of the important role 
of magnetism in cuprate high temperature superconductivity proposed by many 
theoretical models \cite{rice97} and the similar coordination of RuO$_{2}$ and 
CuO$_{2}$ layers in Ru-1212 and Ru-1222, respectively.

Both Ru-1212 and -1222 have a structure \cite{bernhard99,felner97} similar to 
that of the cuprate high 
temperature superconductor GdSr$_{2}$Cu$_{3}$O$_{7}$ = 
CuSr$_{2}$GdCu$_{2}$O$_{7}$ (Gd-123 or Cu-1212) and are 
derived from Cu-1212 by replacing the CuO-chain layer with the RuO$_{2}$ layer. 
For Ru-1222, the oxygen-absent Gd-layer between the two CuO$_{2}$ layers 
in Cu-1212 is further replaced by the double fluorite 
(M$_{0.7}$Ce$_{0.3}$)$_{2}$O$_{2}$ block. The charge reservoir block 
(SrO)(RuO$_{2}$)(SrO) in these compounds has the same atomic arrangement as the 
perovskite SrRuO$_{3}$ \cite{longo68}, which is an itinerant ferromagnet with 
a $T_{m}$ of 
$\sim 160$~K. The detection of a ferromagnetic state in Ru-1212 and -1222 
attributed to the Ru-sublattice is therefore rather natural. What is most 
unexpected and exciting is the detection \cite{bernhard99,felner97} of 
superconductivity in this 
ferromagnetically ordered state below $T_{s} < T_{m}$ with the 
superconductivity assigned to the CuO$_{2}$ layers, similar to the 
cuprate high temperature superconductors. It was proposed \cite{bernhard99} 
that the 
RuO$_{2}$-layer sublattice and the CuO$_{2}$-layer sublattice are only weakly 
coupled and the exchange effect due to the Ru-moment in the CuO$_{2}$ layer is 
small to allow the two ordered states to coexist. A possible p-pairing of the 
superconducting electrons was also mentioned, in which spins of the two electrons 
of the Cooper pair are parallel and the magnetic field effect on its 
superconductivity is less detrimental, similar to that in the superconducting 
ruthenate layered compound Sr$_{2}$RuO$_{4}$ \cite{maeno94}, which has 
a $T_{s} \sim 1$~K.

The ferromagnetic state in Ru-1212 and -1222 has been firmly 
established to be bulk and uniform to a scale of $\sim 20$~\AA\ across 
the samples examined \cite{bernhard99,felner97,felner99}. 
However, the evidence for superconductivity reported to date 
\cite{bernhard99,felner97,bauernfeind95} in these 
compounds has been entirely based on the diamagnetic shift in the magnetic 
susceptibility ($\chi$) measured in the zero-field-cooled (ZFC) mode and the zero 
resistivity ($\rho$) in samples when cooled to below $T_{s}$. However, no 
diamagnetic 
$\chi$-shift in the field-cooled (FC) mode, the conventional signature of a bulk 
superconducting state, has been reported. This raises two interesting 
questions: Is the bulk Meissner state indeed absent in these compounds with bulk 
superconductivity, and if it is, why? We have therefore examined the 
superconducting state of Ru-1212 samples prepared under different conditions. 
We found that there exist no bulk Meissner state and only a 
negligible superconducting condensation energy in the chemically and electrically 
uniform Ru-1212 samples 
studied below $T_{s}$ that show magnetic, electrical, and structural properties 
similar to 
those previously observed \cite{bernhard99,rice97,mclaughlin99}. The absence 
of a bulk 
Meissner state and the superconductivity detected are therefore attributed to 
the possible formation of a sponge-like 
crypto-superconducting fine structure in Ru-1212.

The samples investigated were prepared by thoroughly reacting a mixed 
powder of RuO$_{2}$ (99.95\%), SrCO$_{3}$ (99.99\%), Gd$_{2}$O$_{3}$ (99.99\%) 
and CuO (99.9\%), with a cation composition of Ru:Sr:Gd:Cu = 1:2:1:2, 
in steps similar to those previously reported 
\cite{bernhard99,felner97,bauernfeind95,tallon1}. The mixed 
powder was first heated to 900--960~${}^{\circ}$C in air or in flowing oxygen 
for 10--16~hr to decompose the SrCO$_{3}$. The heated oxide 
was then pulverized, pelletized, and calcined in air, flowing 
nitrogen, or flowing argon at 1010--1020~${}^{\circ}$C for 10--16~hr. 
The process was repeated for some samples. The pellets were then 
powdered, pelletized, and heated in flowing oxygen at 1050--1060~${}^{\circ}$C 
for 10--16~hr. Some of the pellets were subsequently annealed 
in flowing oxygen for 6--7~days at 1055--1070~${}^{\circ}$C. The structure 
was determined by powder X-ray diffraction (XRD), using the 
Rigaku DMAX-IIIB diffractometer; the $\rho$ by the standard four-lead technique, 
employing the 
Linear Research Model LRÐ700 Bridge; and the $\chi$ by the Quantum Design SQUID 
Magnetometer. 

The powder XRD patterns of all of our nominal Ru-1212 samples show 
Ru-1212 as the major phase, with zero to various amounts of 
Sr$_{2}$GdRuO$_{6}$ and SrRuO$_{3}$ as minor magnetic impurities, depending 
on the sample preparation conditions. The prolonged annealing 
at 1055--1070~${}^{\circ}$C does not seem to influence the XRD pattern but 
does increase the sample density. In the present study, we 
report only the results of samples that have undergone prolonged annealing 
and are pure within our XRD resolution of $\sim 3$\%. 
The XRD pattern is in excellent agreement with the 
Rietveld refinement profile for a tetragonal Ru-1212 with space group 
$P4/mmm$ as shown in Fig.~\ref{cvt}. 
The lattice parameters so-determined 
are $a = 3.8375(8)$~\AA\ and $c = 11.560(2)$~\AA, in good agreement with 
values reported previously \cite{mclaughlin99}. Scanning electron 
microscope data show uniform composition across the sample except voids and 
also grains of dimensions $\sim$ 1--5~$\mu$m. 

The $\chi$ measured at 5~Oe during both the FC and ZFC modes is displayed 
in the inset of Fig.~\ref{cvt} 
as a function of temperature. A strong diamagnetic shift starting at $\sim 25$~K 
($T_{s}$) is clearly evidenced in the ZFC-$\chi$ but not in the FC-$\chi$, 
although a ferromagnetic transition appears at $\sim 130$~K in 
both $\chi$'s, similar to the previous report \cite{bernhard99}. The 
diamagnetic shift detected corresponds to a shielding volume fraction of 
$> 100$\% before the demagnetization correction. A drastic drop of 
$\rho$, measured by a current density of $\sim$ 0.5~A/cm$^{2}$, 
starting at $\sim 45$~K and 
reaching zero at $\sim 30$~K ($T_{s}$) 
is observed, similar to that previously reported \cite{bernhard99,mccrone99}. 
$H$ broadens slightly 
the transition and shifts the zero-$\rho$ temperature from $\sim 30$ to 
$\sim 8$~K at 7 T, 
characteristic of a superconducting transition below 
its $T_{s}$.

In a Type II superconductor, a strong flux pinning potential reduces the 
size of the diamagnetic signal of the FC-$\chi$, which, in the 
extreme case, can lead to the apparent 
absence of 
a bulk Meissner effect. We have therefore examined the reversible or equilibrium 
magnetization $M_{r}$ \cite{clem93}, which excludes the flux pinning effect, as 
a function of 
$H$ at temperatures below $T_{s}$. $M_{r}$($H$) is determined as the average 
between the 
magnetizations measured at $H$ during field increase ($M_{+}$) and during 
field decrease ($M_{-}$), \textit{i.e.} 
$M_{r} \equiv [M_{+}(H)]+[M_{-}(H)]/2$, ignoring the surface pinning, 
which is expected to be negligible for cuprate superconductors. 
For a Type-II superconductor, 
$dM_{r}/dH$ is expected to exhibit a large increase from $-1/4\pi$ to 
a positive value as 
$H$ passes over the lower critical field ($H_{c1}$), then to decrease with 
further $H$-increase, and finally to diminish for $H$ $\ge$ the upper 
critical field ($H_{c2}$). For a superconductor with a long-range ferromagnetic 
order, 
such as the case of Ru-1212, the situation may be complicated 
by the possible asymmetrical ferromagnetic hysteresis. Steps 
have been developed to minimize such an effect \cite{xue1}. The ferromagnetism 
will result in an additional positive
contribution to $dM_{r}/dH$, which decreases with increasing field. Both $M_{r}$ 
and 
$dM_{r}/dH$ of Ru-1212 are determined in different field ranges: $\pm 5000$~Oe 
in steps 
of 100~Oe, $\pm 300$~Oe in steps of 10~Oe, $\pm 100$~Oe in steps 
of 1~Oe, and $\pm 30$~Oe in steps of 0.25~Oe. The 
results for $\pm 300$~Oe are shown in Fig.~\ref{rvt}. $M_{r}$ does not show 
any deviation from the almost linear magnetic $H$-dependence nor does 
$dM_{r}/dH$ display the large increase expected of a Meissner state 
as $H$ passes over $H_{c1}$, even for $H$ as small as $\pm 0.25$~Oe. The 
results unambiguously show the absence of a bulk Meissner state 
in our Ru-1212 samples to $H=0$. However, the initial slight 
increase of $dM_{r}/dH$ with $H$ allows us to estimate that no more than 
a few percent of a bulk Meissner state exists in the samples 
investigated. 

A spontaneous vortex lattice (SVL) has been proposed to form in a 
ferromagnetic superconductor \cite{ferrel79} or in a superconducting ferromagnet 
\cite{sonin98}, 
provided that the internal magnetic field ($4\pi M_{m}$) associated with the 
magnetic moment ($M_{m}$) is greater than $H_{c1}$ but smaller than $H_{c2}$. 
The formation of the SVL can substantially reduce the 
size of the Meissner signal, similar to that observed in Ru-1212. However, the 
condensation energy per volume ($F_{s}$) associated with the superconductivity 
in Ru-1212 may be 
estimated as $$F_{s} = 
(1/4\pi)\int^{H_{c2}}_{0}M_{s}dH,$$ which can still be large, 
since $H_{c2} > 7$~T $\gg 4\pi M_{m} \sim$ 200--700~Oe, where $M_{s}$ is 
the reversible moment associated with superconductivity. In Ru-1212, 
$M_{r} = M_{s} + M_{m}$. To 
determine $M_{s}$, we have measured $M_{r}$'s of the superconducting Ru-1212 
and the non-superconducting Ru-1212Zn 
[RuSr$_{2}$GdCu$_{1.94}$Zn$_{0.06}$O$_{8}$] 
at 5~K ($< T_{s}$ of Ru-1212) and 50 K ($> T_{s}$) up to 600~Oe. Both Ru-1212 and 
Ru-1212Zn exhibit a 
similar $T_{m}$ of $130\pm2$ K. $M_{s}$ is then obtained as 
[$M_{rRu-1212} 
\hbox{ (5~K)} - 
M_{rRu-1212Zn} \hbox{ (5~K)}] - [M_{rRu-1212} 
\hbox{ (50~K)} - 
M_{rRu-1212Zn}$ (50~K)] and is shown in Fig.~\ref{mr}. The second term in the 
parentheses is used to eliminate the change in magnetic 
properties induced by the Zn-substitution. The $M_{r}$ ($=M_{s}$) of a 
YBa$_{2}$Cu$_{3}$O$_{6.6}$ (YBCO) sample with a bulk Meissner state and a 
$T_{s}$ = 40~K, was 
also measured at 5~K and is displayed with previous high field data 
\cite{daumling 91} in 
the same figure for comparison. It is evidenced that $F_{s}$ of Ru-1212 at 
5~K is very small and lies within the resolution of the measurements 
of $\sim$ a few \% of that of YBCO at the same temperature. This is 
also in agreement with the absence of a bulk Meissner state detected 
by us. It was 
shown \cite{daumling 91} previously that the condensation energy so determined 
is in 
agreement with that measured calorimetrically. Therefore, a 
specific heat anomaly ($\Delta C_{p}$) associated with $F_{s}$ at $T_{s}$ of 
only a few \% that for a bulk superconductor with a similar $T_{s}$ is expected. 
This is in contrast to the $\Delta C_{p}$ at $T_{s}$ recently 
reported to be of the same 
size as that for Bi-1212 \cite{tallon2}. Adding to the puzzle is the enhancement 
of the peak-temperature of $\Delta C_{p}$ by field detected \cite{xue2}, 
reminiscent of an 
antiferromagnetic 
origin of 
the $\Delta C_{p}$. It is known that the antiferromagnet Sr$_{2}$GdRuO$_{6}$ 
appears often 
in Ru-1212 and has a Ne\'el temperature $\sim$ 30's~K. A possible small 
inclusion ($<$ 1\%) of such an antiferromagnet in the Ru-1212 sample can 
account for the observation. Repeating the $C_{p}$-measurement on Ru-1212 
samples with $<$ 1\% Sr$_{2}$GdRuO$_{6}$ is warranted, although the measurement 
is far from being trivial, as already demonstrated \cite{tallon2}, because of the strong 
magnetic contribution from the sample.

As described earlier, the samples investigated are structurally pure 
to the XRD resolution of $\sim$ 3\% and chemically uniform to the 
microprobe resolution of 1--2 $\mu$m across the sample. Recently, both the 
$ac$ and $dc$ magnetizations of Ru-1212 samples, all powdered 
sequentially from the same bulk piece of ours to particles of sizes varying 
from 10 to 
800 $\mu$m with grain sizes of 1--5 $\mu$m, 
were measured in fields 
ranging from $10^{-4}$ to $10^{2}$~Oe \cite{xue2}. It was found that the 
hysteric part of 
the $dc$ $\chi$ and the imaginary part of the $ac$ $\chi$ scale linearly with 
the size of the particles ($d$) but the real part of the $ac$ $\chi$ decreases 
with 
$d$ systematically and drastically. The former, which is a measure of 
critical current density and of the electrical connectivity throughout 
the particles, suggests that the Ru-1212 
sample examined is electrically uniform to $\sim$ 10 $\mu$m and the 
latter, 
which is a measure of screening, implies a penetration depth as large as 
30--50 $\mu$m, suggesting a very small effective carrier density, 
which is consistent with the negligible condensation energy observed 
by us.

To account for the absence of a bulk Meissner state, the appearance of 
negligible condensation energy, and the presence of a large effective 
penetration depth in a structurally pure ($\sim$ 3\%), chemically uniform 
(to $\sim$ 1--2 
$\mu$m), and electrically homogeneous (to 10 $\mu$m) Ru-1212, we propose 
that there may exist a sponge-like superconducting network distributed 
uniformly across the Ru-1212 sample. In other words, the 
superconducting order may have modified itself with a fine structure 
(crypto-superconductivity) in order to fit the ferromagnetic state in 
the superconducting ferromagnet Ru-1212, in a way very similar to the 
proposed crypto-ferromagnetism in a ferromagnetic superconductor. One 
possible scenario is for crypto-superconductivity to exist in the 
ferromagnet-domain boundaries where the magnetic field can be smaller 
than $H_{c2}$ and the magnetic scattering is suppressed. The possible 
formation of these fine ferromagnet-domains in the chemically uniform 
itinerant ferromagnet Ru-1212, driven by the electromagnetic 
interaction, appears to be not unreasonable in view of what has been 
observed recently in the closely related highly correlated colossal 
magnetoresistance manganites \cite{goodenough97}.

In conclusion, we have studied the structural, electrical, 
and magnetic properties of Ru-1212 samples prepared under 
different conditions previously reported. The reversible 
magnetization data clearly show the absence of a bulk Meissner state 
in Ru-1212 samples that are structurally pure, chemically 
homogeneous, and electrically uniform. By examining the difference 
between the reversible magnetization of the superconducting Ru-1212 
and the non-superconducting Ru-1212Zn, we conclude that the 
condensation energy associated with the superconducting state is very 
small, not more than a few \% of a bulk cuprate superconductor with a 
similar $T_{s}$. To account for the observations, we propose the possible 
formation of crypto-superconductivity in the superconducting 
ferromagnet, similar to the formation of crypto-ferromagnetism in a 
ferromagnetic superconductor previously suggested.

Work done in Houston is supported in part by NSF Grant No.~DMR-9804325, 
the T.~L.~L.~Temple Foundation, the John J.~and Rebecca Moores Endowment, 
and the State of Texas through the Texas Center for Superconductivity at 
the University of Houston; and at Lawrence Berkeley Laboratory 
by the Director, Office of Energy Research, Office of Basic 
Energy Sciences, Division of Materials Sciences of the U.S. 
Department of Energy under Contract No.~DE-AC03-76SF00098.

\begin{figure}
\caption{XRD pattern of Ru-1212: dots --- data and solid line --- 
Rietveld refinement profile. The inset shows both the FC-$\chi$ 
($\bigtriangledown$) 
and the ZFC-$\chi$ ($\bigcirc$) measured at 5~Oe.}
\label{cvt}
\end{figure}

\begin{figure}
\caption{$M_{r}$ and $dM_{r}/dH$ \textit{vs.}~$H$ of Ru-1212 at 2~K.}
\label{rvt}
\end{figure}

\begin{figure}
\caption{$M_{s}$ \textit{vs.}~$H$ for Ru-1212 ($\bullet$) and for 
YBCO ($\Box$, 
$\bigtriangleup$) at 5~K. Refer to the text for the detemination of $M_{s}$. The 
inset shows $M_{r}$ \textit{vs.}~$H$ for superconducting Ru-1212 
($\bullet$) 
and non-superconducting Ru-1212Zn ($\Box$) at 5~K.}
\label{mr}
\end{figure}

\end{document}